\PassOptionsToPackage{unicode}{hyperref}
\PassOptionsToPackage{hyphens}{url}
\documentclass[
  preprint,
  12pt]{elsarticle}
\usepackage{amsmath,amssymb}
\usepackage{lmodern}
\usepackage{iftex}
\ifPDFTeX
  \usepackage[T1]{fontenc}
  \usepackage[utf8]{inputenc}
  \usepackage{textcomp} 
\else 
  \usepackage{unicode-math}
  \defaultfontfeatures{Scale=MatchLowercase}
  \defaultfontfeatures[\rmfamily]{Ligatures=TeX,Scale=1}
\fi
\IfFileExists{upquote.sty}{\usepackage{upquote}}{}
\IfFileExists{microtype.sty}{
  \usepackage[]{microtype}
  \UseMicrotypeSet[protrusion]{basicmath} 
}{}
\usepackage{xcolor}
\IfFileExists{xurl.sty}{\usepackage{xurl}}{} 
\IfFileExists{bookmark.sty}{\usepackage{bookmark}}{\usepackage{hyperref}}
\hypersetup{
  hidelinks,
  pdfcreator={LaTeX via pandoc}}
\usepackage{graphicx}
\makeatletter
\def\maxwidth{\ifdim\Gin@nat@width>\linewidth\linewidth\else\Gin@nat@width\fi}
\def\maxheight{\ifdim\Gin@nat@height>\textheight\textheight\else\Gin@nat@height\fi}
\makeatother
\setkeys{Gin}{width=\maxwidth,height=\maxheight,keepaspectratio}
\makeatletter
\def\fps@figure{htbp}
\makeatother
\usepackage{bm}
\usepackage{braket}
\usepackage[superscript]{cite}
\newcounter{bla}

\journal{Computer Physics Communications}

\usepackage[capitalise]{cleveref}

\creflabelformat{equation}{#2#1#3}

\crefname{table}{Table}{Tables}
\ifLuaTeX
  \usepackage{selnolig}  
\fi
\newlength{\cslhangindent}
\newlength{\csllabelwidth}
\newenvironment{CSLReferences}[2] 
 {
  \setlength{\parindent}{0pt}
  \ifodd #1 \everypar{\setlength{\hangindent}{\cslhangindent}}\ignorespaces\fi
  \ifnum #2 > 0
  \setlength{\parskip}{#2\baselineskip}
  \fi
 }%
 {}
\usepackage{calc}

\newcommand{\CSLLeftMargin}[1]{\parbox[t]{\csllabelwidth}{#1}}
\newcommand{\CSLRightInline}[1]{\parbox[t]{\linewidth - \csllabelwidth}{#1}\break}

\begin{document}

\begin{frontmatter}

\title{First-principles LCPAO Approach for Insulators under Finite Electric Fields}
\author[a]{Naoya Yamaguchi\corref{author1}}
\author[a]{Fumiyuki Ishii\corref{author2}}
\cortext[author1] {\textit{E-mail address:} n-yamaguchi@cphys.s.kanazawa-u.ac.jp}
\cortext[author2] {\textit{E-mail address:} ishii@cphys.s.kanazawa-u.ac.jp}
\address[a]{Nanomaterials Research Institute (NanoMaRi), Kanazawa University, Kakuma-machi, Kanazawa, 920-1192, Japan}

\begin{abstract}
    We propose a linear-combination-of-pseudo-atomic-orbitals scheme for a finite electric field method based on the modern theory of polarization. We derive the matrix elements of the effective potential for the field and the corresponding terms of the forces on atoms.
    In addition, we successfully evaluated the dielectric constants and Born effective charges of typical semiconducting and insulating materials.
    Our formalism will aid in the study of materials under electric fields.
\end{abstract}


\makeatletter
\def\ext@figure{}
\def\ext@table{}
\makeatother

\end{frontmatter}

\hypertarget{introduction}{%
\section{Introduction}\label{introduction}}

The effects of electric fields on insulators are of significant research interest, and have been investigated experimentally and theoretically for several applications. In theoretical works based
on quantum mechanics, dielectric properties including responses for
electric fields have been investigated intensively since the past fifty years~{[}1--3{]}. In the modern theory of polarization (MTP)~{[}4{]}, first-principles calculations can be used on insulating solids to determine the electric polarization as a well-defined
bulk quantity with Berry phases. In addition, polarization and Berry
phases can be leveraged to probe the electronic structure of solids under
electric fields~{[}5{]}.

MTP addresses the difficulty in describing electric polarization in bulk
insulators with Berry phases obtained from overlap matrices between the
periodic parts of Bloch orbitals at a k-point and the next k-point,
instead of explicit evaluation of the position operator. The difficulty
arises from the periodic boundary condition applied in electronic
structure calculations based on the density functional theory (DFT), and
it cannot be determined how electrons cross the cell boundary. This is because the existence of electrons is described by continuous electronic densities. Considering electric fields in the DFT framework, the boundary condition leads to another difficulty in that the periodicity can yield unphysical fields.
Additional potentials for finite electric fields include terms arising from the
product of the electric field and position operator
\(\bm{E}\cdot\bm{r}\), but it is not acceptable when there are electrons
at the boundary (i.e.~saw-tooth potential in bulk systems).
The position can be substituted with the polarization \(\bm{P}\) to
consider the potential for electric fields in bulk insulators with
electric-field-polarized Wannier functions~{[}6{]}, because MTP shows that polarization is independent of the boundary, except for
arbitrary modules of polarization. The first DFT
calculations using \(\bm{E}\cdot\bm{P}\) were reported in 1998, but
large localization regions were required to get converged
field-polarized Wannier functions~{[}7{]}. In 2002, practical DFT
schemes for \(\bm{E}\cdot\bm{P}\) were proposed~{[}5{]}, and they were
implemented within the plane wave method~{[}5{]} or Car--Parrinello
method~{[}8{]}. The electronic structures of
III-V or II-VI semiconductors were successfully evaluated under finite electric fields, and
dielectric properties such as dielectric constants and the Born
effective charges were also calculated. Such a scheme was also implemented in a projector
augmented wave framework~{[}9{]}.

For realistic conditions, large scale calculations are generally required, and
localized orbitals such as the linear combination of atomic orbitals
(LCAO) method are appropriate in terms of scalability. Interesting
models to consider the electric field effect on vacancies, impurities, and
interfacial systems may require hundreds or thousands of atoms.
Moreover, MTP for the LCAO scheme demands a conversion from Bloch orbitals to
atomic orbitals~{[}10{]}. Generally, atomic orbitals do not comprise the
complete basis set, and therefore, Pulay's correction is required.
These requirements make implementation complicated, which is a
difficulty in the LCAO scheme. In the time-dependent
\text{schr\"odinger} equation, a time-evolving tight-binding scheme for
finite electric fields was proposed based on MTP~{[}11{]}. It was
implemented in the time-dependent DFT framework of the SIESTA code, a
linear combination of pseudo atomic orbitals (LCPAO) first-principle code~{[}12,13{]} (the implementation details have
not been reported). However, with regard to
time-independent schemes, only a density matrix approach in the
LCAO scheme for electric field effects has been reported thus far~{[}14{]}. The LCAO
implementation of effective potential and the forces on atoms for finite
electric fields practically applicable to ordinal DFT is expected to
expand the application range to consider even large scale
or realistic models (e.g.~vacancies in the diamond nitrogen-vacancy center) or
complicated magnetic systems (e.g.~spin spirals with the spin-orbit
interaction (SOI)).

In this study, we developed a first LCAO implementation of the finite
electric field scheme including the forces on atoms compatible with
self-consistent field (SCF) loops in the DFT framework based on MTP. In addition, we evaluated electronic structures under uniform electric fields. Moreover, we calculated the dielectric properties of insulators and semiconductors: dielectric constants and Born effective charges of III-V and II-VI
semiconductors and group IV insulating materials. The calculated
dielectric properties agree with previously reported computational values.
The LCAO implementation should serve as a powerful tool to investigate
electric field effects on large scale systems.

\hypertarget{theory}{%
\section{Theory}\label{theory}}

\hypertarget{lcpao-method-and-modern-theory-of-electric-polarization}{%
\subsection{LCPAO method and modern theory of electric
polarization}\label{lcpao-method-and-modern-theory-of-electric-polarization}}

The LCPAO method uses basis sets localized on atoms, and the basis functions are based on atomic orbitals~{[}10{]}. The
required number of basis functions is significantly smaller than that in the plane wave
method, which enables solving eigenvalue problems by direct
diagonalization. The electronic contribution to electric polarization is
described via the Berry phase using Bloch orbitals. While
Bloch orbitals can be expressed easily by plane wave basis sets, using localized basis sets of PAOs Bloch orbitals in the
LCPAO method are
given as 
\begin{equation}
\braket{\bm{r}|\psi_{\mu}^{(\sigma\bm{k})}}
=\frac{1}{\sqrt{N}}\sum_{n=0}^{N-1}e^{\mathrm{i}\bm{k}\cdot\bm{R}_n}\sum_{i\alpha}c^{(\sigma\bm{k})}_{i\alpha, \mu}\braket{\bm{r}|\phi_{i\alpha}^{\bm{R}_n}},
\label{eq:LCPAO}
\end{equation} 
where \(\bm{r}\) is the position, \(N\)
is the number of cells considered in calculations, ``\(\mathrm{i}\)'' is
the imaginary unit, \(\bm{k}\) is a wave number, \(\bm{R}_n\) is the
lattice vector for cell \(n\), \(c^{(\sigma\bm{k})}_{i\alpha, \mu}\) is
an LCPAO coefficient connecting a PAO for orbital \(\alpha\) belonging to
atom \(i\) with a Bloch orbital of spin \(\sigma\) and state \(\mu\),
\(\tau_i\) is the position of an atom \(i\), and
\(\phi_{i\alpha}^{\bm{R}}\) is a PAO for orbital \(\alpha\) belonging to
atom \(i\) in a cell moved by \(\bm{R}\) from the original cell, and it is expanded around $\tau_i$, i.e.~$\braket{\bm{r}|\phi_{i\alpha}^{\bm{R}}}=\phi_{i\alpha}(\bm{r}-\bm{\tau}_i-\bm{R})$. For the
LCPAO methods with the Hamiltonian
\(\mathrm{H}^{(\sigma\bm{k})}_{i\alpha, j\beta}=\sum_{n=0}^{N-1}e^{\mathrm{i}\bm{k}\cdot\bm{R}_n}\braket{\phi_{i\alpha}^{\bm{0}}|H_\sigma|\phi_{j\beta}^{\bm{R}_n}}\)
and the overlap matrix
\(\mathrm{S}^{(\bm{k})}_{i\alpha, j\beta}=\sum_{n=0}^{N-1}e^{\mathrm{i}\bm{k}\cdot\bm{R}_n}\braket{\phi_{i\alpha}^{\bm0}|\phi_{j\beta}^{\bm{R}_n}}\),
the Kohn--Sham equation is given as
\begin{equation}
\mathrm{H}^{(\sigma\bm{k})}c^{(\sigma\bm{k})}=\epsilon^{(\sigma\bm{k})}\mathrm{S}^{(\bm{k})}c^{(\sigma\bm{k})},
\label{eq:KS}
\end{equation} 
and can be solved by direct diagonalization
to obtain the eigenvalues of energies \(\epsilon^{(\sigma\bm{k})}\). According
to MTP~{[}4{]}, the electronic contribution to electric polarization
\(\bm{P}_e\) in an insulating system with periodic cells is obtained
from \(\bm{G}_a\cdot\bm{P}_e\), \(\bm{G}_b\cdot\bm{P}_e\), and
\(\bm{G}_c\cdot\bm{P}_e\), which can be expressed by the Berry phase
as 
\begin{equation}
\bm{G}_a\cdot\bm{P}_e
=-\frac{ef}{\Omega N_bN_c}\sum_{\sigma}\sum_{I_b=0}^{N_b-1}\sum_{I_c=0}^{N_c-1}\mathrm{Im}\ln\det\prod_{I_a=0}^{N_a-1}M^{(\sigma, I_a, I_b, I_c)}_a,
\label{eq:MTP}
\end{equation} 
where \(f\) is a weighting coefficient of
the spin degeneracy per state, \(\bm{G}_a\), \(\bm{G}_b\) and
\(\bm{G}_c\) are the reciprocal lattice vectors for the cell vectors
\(\bm{\mathrm{v}}_a\), \(\bm{\mathrm{v}}_b\), and \(\bm{\mathrm{v}}_c\),
respectively, \(e\) is the elementary charge, \(\Omega\) is the cell
volume, \(N_a\), \(N_b\) and \(N_c\) are the numbers of k-points to
discretize the first Brillouin zone along \(\bm{G}_a\), \(\bm{G}_b\) and
\(\bm{G}_c\), respectively,
\(M^{(\sigma, I_a, I_b, I_c)}_{a\mu\nu}=\Braket{u^{(\sigma\bm{k}(I_a, I_b, I_c))}_\mu|u^{(\sigma\bm{k}(I_a+1, I_b, I_c))}_\nu}\)
is an overlap matrix of the periodic parts \(u^{\sigma\bm{k}}_\mu\) of
occupied Bloch orbitals between adjacent k-points, and
\(\bm{k}(I_a, I_b, I_c)=(I_a/N_a)\bm{G}_a+(I_b/N_b)\bm{G}_b+(I_c/N_c)\bm{G}_c\).
The manner of expressing electric polarization by the Berry phase has been
successful, since it provides evaluations based on well-defined
quantities instead of direct evaluation of the position \(\bm{r}\).
\(M^{(\sigma, I_a, I_b, I_c)}_a\) can be calculated by
\begin{equation}
M^{(\sigma, I_a, I_b, I_c)}_{a\mu\nu}
=\left(c^{(\sigma\bm{k}(I_a, I_b, I_c))\dag}\mathrm{T}^{(\bm{k}(I_a+1, I_b, I_c))}_ac^{(\sigma\bm{k}(I_a+1, I_b, I_c))}\right)_{\mu\nu},
\label{eq:M_CTC}
\end{equation} 
where we introduced
\(\mathrm{T}^{(\bm{k})}_{ai\alpha, j\beta}=\sum_{n=0}^{N-1}e^{\mathrm{i}\bm{k}\cdot\bm{R}_n}\Braket{\phi_{i\alpha}^{\bm0}|e^{-\mathrm{i}\frac{\bm{G}_a}{N_a}\cdot\bm{r}}|\phi_{j\beta}^{\bm{R}_n}}\),
and \(\mu\) and \(\nu\) run over occupied states. Although
\(M^{(\sigma, \bm{k}(I_a, I_b, I_c), \bm{k}(I_a+1, I_b, I_c))}\) itself
is gauge-dependent, the dependence vanishes when considering the sum over the
first Brillouin zone, and the electric polarization \(P_e\) can be evaluated. By
introducing \(\mathrm{T}^{(\bm{k})}_{ai\alpha, j\beta}\), the known
LCPAO formalism of \(\Braket{u^{(\sigma\bm{k}(I_a, I_b, I_c))}_\mu|u^{(\sigma\bm{k}(I_a+1, I_b, I_c))}_\nu}\)~{[}10{]} becomes a matrix form of
\cref{eq:M_CTC}, which is most suitable for computation in terms of
efficiency and simplicity, especially for iterative calculations such as
SCF calculations. This is because it suffices to prepare
\(\mathrm{T}^{(\bm{k})}_{ai\alpha, j\beta}\) once at the beginning of
each SCF loop.

\hypertarget{effective-potential-for-finite-electric-fields}{%
\subsection{Effective potential for finite electric
fields}\label{effective-potential-for-finite-electric-fields}}

We start with an expression of the electric enthalpy functional
\(\mathcal{F}\), which is the total energy for an insulating system under a
finite electric field:
\(\mathcal{F}[\bm{E}]=\mathcal{E}_{\mathrm{KS}}-\Omega\bm{E}\cdot\bm{P}\)
~{[}6{]}. Here, \(\bm{E}\) is the electric field,
\(\mathcal{E}_{\mathrm{KS}}\) is the ordinary counterpart of Kohn--Sham
total energy described by field-polarized orbitals, and the polarization
\(\bm{P}\) comprises the ionic and electronic contributions:
\(\bm{P}=\bm{P}_I+\bm{P}_e\). The solution that minimizes
\(\mathcal{F}[\bm{E}]\) is desirable, such that diagonalization
can be used to get solutions for insulators under electric fields in the same manner as in
variational approaches to the ordinary Kohn--Sham equation. Here, considering an additional term of the effective potential
\(V=-\Omega\bm{E}\cdot\bm{P}\) is required. In the plane wave
method, Souza \textit{et al.} suggested the conjugate-gradient method
using a gradient
\(\delta\mathcal{F}/\delta\bra{u^{(\sigma\bm{k})}_\mu}\)~{[}5{]}. For
the LCPAO method, we propose direct diagonalization by adding
matrix elements of the effective potential into those of the ordinary
Kohn--Sham Hamiltonian \(H_0\). The matrix representation
of the effective potential \(A\) can be extracted from the relation
\(V=\sum_{\sigma\bm{k}}\mathrm{tr}(c^{(\sigma\bm{k})\dag}A^{(\sigma\bm{k})}c^{(\sigma\bm{k})})\).
Considering the derivative of \(V\) with respect to the LCPAO coefficients,
we get 
\begin{equation}
\frac{\partial V}{\partial c^{(\sigma\bm{k})*}_{i\alpha, \mu}}
=(A^{(\sigma\bm{k})}c^{(\sigma\bm{k})})_{i\alpha, \mu}.
\label{eq:dVdc1}
\end{equation} 
Moreover, we expand \cref{eq:dVdc1} with the
cell vectors \(\bm{\mathrm{v}}_a\), \(\bm{\mathrm{v}}_b\), and
\(\bm{\mathrm{v}}_c\), and obtain 
\begin{equation}
\frac{\partial V}{\partial c^{(\sigma\bm{k})*}_{i\alpha, \mu}}
=-\Omega\bm{E}\cdot\frac{\partial\bm{P}}{\partial c^{(\sigma\bm{k})*}_{i\alpha, \mu}}
=-\Omega\bm{E}\cdot\frac{\partial\bm{P}_e}{\partial c^{(\sigma\bm{k})*}_{i\alpha, \mu}}
=-\frac{\Omega}{2\pi}\sum_{\lambda=a, b, c}(\bm{E}\cdot\bm{\mathrm{v}}_\lambda)\frac\partial{\partial c^{(\sigma\bm{k})*}_{i\alpha, \mu}}(\bm{G}_\lambda\cdot\bm{P}_e).
\label{eq:dVdc2}
\end{equation} 
Here, 
\begin{equation}
\frac\partial{\partial c^{(\sigma\bm{k})*}_{i\alpha, \mu}}(\bm{G_a}\cdot\bm{P}_e)
=-\frac{ef}\Omega\left(\left(\frac{O^{(\sigma\bm{k})}_a-O^{(\sigma\bm{k})\dag}_a}{2i}\right)c^{(\sigma\bm{k})}\right)_{i\alpha, \mu},
\label{eq:dGPdc}
\end{equation} 
where
\(O^{(\sigma\bm{k}(I_a, I_b, I_c))}_a=\) 
\small 
\begin{equation}
\frac{1}{N_bN_c}\mathrm{T}^{\bm{k}(I_a+1, I_b, I_c)}_ac^{(\sigma\bm{k}(I_a+1, I_b, I_c))}\left(M^{(\sigma, I_a, I_b, I_c)}_a\right)^{-1}\left(M^{(\sigma, I_a-1, I_b, I_c)}_a\right)^{-1}c^{(\sigma\bm{k}(I_a-1, I_b, I_c))\dag}\mathrm{T}^{\bm{k}(I_a, I_b, I_c)}_a.
\label{eq:O}
\end{equation} 
\normalsize Then, we get 
\begin{equation}
\frac{\partial V}{\partial c^{(\sigma\bm{k})*}_{i\alpha, \mu}}
=\left(\left(\frac{ef}{2\pi}\sum_{\lambda=a, b, c}(\bm{E}\cdot\bm{\mathrm{v}}_\lambda)\frac{O^{(\sigma\bm{k})}_\lambda-O^{(\sigma\bm{k})\dag}_\lambda}{2i}\right)c^{(\sigma\bm{k})}\right)_{i\alpha, \mu},
\label{eq:dVdc3}
\end{equation} 
Therefore, the matrix elements of \(A\)
are given as 
\begin{equation}
A^{(\sigma\bm{k})}_{i\alpha, j\beta}
=\left(\frac{ef}{2\pi}\sum_{\lambda=a, b, c}(\bm{E}\cdot\bm{\mathrm{v}}_\lambda)\frac{O^{(\sigma\bm{k})}_\lambda-O^{(\sigma\bm{k})\dag}_\lambda}{2i}\right)_{i\alpha, j\beta}.
\label{eq:A}
\end{equation} 
Finally, in the case of a finite electric
field, the Hamiltonian
\(\mathrm{H}^{(\sigma\bm{k})}=H_0^{(\sigma\bm{k})}+A^{(\sigma\bm{k})}\)
can be evaluated to obtain a solution through \cref{eq:KS}. In addition, this formalism is
compatible with non-collinear spin density functionals, and is applicable
to cases considering the SOI.

\hypertarget{forces-under-finite-electric-fields}{%
\subsection{Forces under finite electric
fields}\label{forces-under-finite-electric-fields}}

For the forces on atoms induced by the effective potential, the ionic
contribution is
\(\Omega\bm{E}\cdot\frac{\partial\bm{P}_I}{\partial\bm{\tau}_i}=Z_ie\bm{E}\),
where \(Z_i\) is the core charge of atom \(i\), and the electronic
contribution in the case of the complete basis set is taken into
account without additional terms, owing to the Hellmann--Feynman theorem.
In the LCAO representation, the additional terms, i.e.~Pulay forces,
are necessary, so that the electronic contribution is 
\begin{equation}
\Omega\bm{E}\cdot\frac{\partial\textbf{P}_e}{\partial\bm{\tau}_i}
=\frac{ef}{2\pi}\sum_{\sigma\bm{k}}\sum_{\lambda=a, b, c}(\bm{E}\cdot\bm{\mathrm{v}}_\lambda)\mathrm{Im}\mathrm{tr}\left(\left(\frac{\partial c^\dag}{\partial\bm{\tau}_i}\textrm{T}c+c^\dag\frac{\partial \textrm{T}}{\partial\bm{\tau}_i}c+c^\dag\textrm{T}\frac{\partial c}{\partial\bm{\tau}_i}\right)M^{-1}\right),
\label{eq:force}
\end{equation} 
where
\(c^\dag=c^{(\sigma\bm{k}(I_a, I_b, I_c))\dag}\),
\(\mathrm{T}=\mathrm{T}^{(\bm{k}(I_a+1, I_b, I_c))}_a\),
\(c=c^{(\sigma\bm{k}(I_a+1, I_b, I_c))}\), \(M=c^\dag\mathrm{T}c\).
Since
\(\mathcal{F}=\sum_{\sigma\bm{k}}\mathrm{tr}(c^{(\sigma\bm{k})\dag}\mathrm{H}^{(\sigma\bm{k})}c^{(\sigma\bm{k})})\),
the terms of
\(\frac{\partial c^\dag}{\partial\bm{\tau}_i}\textrm{T}c+c^\dag\textrm{T}\frac{\partial c}{\partial\bm{\tau}_i}\)
are part of
\(\frac{\partial c^\dag}{\partial\bm{\tau}_i}\textrm{H}c+c^\dag\textrm{H}\frac{\partial c}{\partial\bm{\tau}_i}\),
calculated from the energy density matrix
\(\varepsilon_{i\alpha, j\beta}^{(\sigma\bm{R}_n)}=\frac{\Omega}{2\pi}\sum_{\mu}^{\mathrm{occ.}}\sum_{\bm{k}}e^{\mathrm{i}\bm{k}\cdot\bm{R}_n}\epsilon^{(\sigma\bm{k})}_{\mu}c^{(\sigma\bm{k})*}_{i\alpha, \mu}c^{(\sigma\bm{k})}_{j\beta, \nu}\).
The derivative of the overlap matrix is
\(\mathrm{s}_{i\alpha, j\beta}^{(\bm{R}_n)}=\braket{\phi_{i\alpha}^{\bm0}|\phi_{j\beta}^{\bm{R}_n}}\)~{[}10{]}.
After SCF convergence, only the
\(c^\dag\frac{\partial \textrm{T}}{\partial\bm{\tau}_i}c\) term needs to be
calculated explicitly, while evaluation of other terms in the
electronic contribution is replaced with that of the
\(\frac{\partial c^\dag}{\partial\bm{\tau}_i}\textrm{H}c+c^\dag\textrm{H}\frac{\partial c}{\partial\bm{\tau}_i}\)
This is realized using \(\varepsilon^{(\sigma\bm{R}_n)}\) determined
after SCF convergence with the derivative of \(s^{(\bm{R}_n)}\).

\hypertarget{computational-details}{%
\section{Computational details}\label{computational-details}}

We implemented the electric field code based on our LCPAO formalism on the OpenMX code~{[}15--17{]} (\texttt{http://www.openmx-square.org}),
which performs first-principles calculations based on DFT within the
local density approximation (LDA)~{[}18{]} or generalized gradient
approximation (GGA)~{[}19{]}. The norm-conserving pseudopotentials we
used included the 2s and 2p electrons for C and O as valence electrons,
while the 2p and 3s for Mg; 3s and 3p for Al, Si, P and S; 3d
and 4s for Zn; 3d, 4s and 4p for Ga, Ge and As; 4s and 4p for
Se; 4d and 5s for Cd; and 4d, 5s and 5p for In and Sb were also considered. We used a
\(32\times32\times32\) regular k-point grid and real space grids
corresponding to energy cutoffs larger than 300 Ry to obtain the converged
results of the dielectric constants and Born
effective charges. We also used the following \emph{standard} and
\emph{precise} PAO basis sets of\(Xr\)-s\(n_s\)p\(n_p\)d\(n_d\)f\(n_f\), where \(X\), \(r\) and \(n_s\),
\(n_p\), \(n_d\) and \(n_f\) denote the element and cutoff radius in the
unit of Bohr and numbers of s-, p-, d- and f-orbital sets, respectively:
for the \emph{standard} one, C6.0-s2p2d1, O6.0-s2p2d1, Mg9.0-s3p2d1,
Al7.0-s2p2d1, Si7.0-s2p2d1, P7.0-s2p2d1f1, S7.0-s2p2d1f1, Zn6.0S-s3p2d1,
Ga7.0-s3p2d2, Ge7.0-s3p2d2, As7.0-s3p2d2, Se7.0-s3p2d2, Cd7.0-s3p2d2,
In7.0-s3p2d2 and Sb7.0-s3p2d2; for the \emph{precise} one, C6.0-s3p2d2,
O6.0-s3p2d2, Mg9.0-s3p2d2, Al7.0-s3p2d2, Si7.0-s3p3d2, P7.0-s3p2d2f1,
S7.0-s3p2d2f1, Zn6.0S-s3p2d2f1, Ga7.0-s3p2d2f1, Ge7.0-s3p2d2f1,
As7.0-s3p2d2f1, Se7.0-s3p2d2f1, Cd7.0-s3p2d2f1, In7.0-s3p2d2f1 and
Sb7.0-s3p2d2f1.

We used computational models where the unit cell is the
primitive cell of face centered cubic or zinc-blende structures with two
atoms for computation. We applied an electric field of 0.1 GV/m along
the a-axis of the conventional cubic cell including eight atoms. We
optimized the lattice constants with the \emph{precise} PAOs under
the LDA, and w neglected the changes due to an external electric
field. The optimized lattice constants were \(a=2.4989~\AA\) for C,
\(a=3.8219~\AA\) for Si, \(a=3.8471~\AA\) for AlP, \(a=3.9845~\AA\) for
AlAs, \(a=3.8367~\AA\) for GaP, \(a=3.9998~\AA\) for GaAs,
\(a=4.3184~\AA\) for AlSb, \(a=4.1413~\AA\) for InP, \(a=3.7529~\AA\)
for ZnS, \(a=3.9356~\AA\) for ZnSe, \(a=4.2526~\AA\) for ZnTe,
\(a=4.2446~\AA\) for CdSe, \(a=4.5433~\AA\) for CdTe, and
\(a=2.9410~\AA\) for MgO. The dielectric constant tensor \(\varepsilon\)
was evaluated from
\begin{equation}
\varepsilon_{\alpha\beta}=\delta_{\alpha\beta}+\frac{1}{\epsilon_0}\frac{\partial P_\alpha}{\partial E_\beta}\quad (\alpha,\beta=x,y,z),\label{eq:epsilon}
\end{equation}
where \(\epsilon_0\) and \(\delta_{\alpha\beta}\) are the electric
constant and Kronecker delta, respectively. When the electric
polarization induced by the electric field \(\Delta\bm{P}\) is given
with and without atomic relaxation, \(\varepsilon_{\alpha\beta}\) is
equal to the static and electronic dielectric constants
\(\varepsilon_{\mathrm{s}\alpha\beta},\varepsilon_{\infty\alpha\beta}\),
respectively, and the following can be derived
\begin{equation}\varepsilon_{\infty xx}=1+\frac{1}{\epsilon_0}\left.\frac{\Delta P_\alpha}{E_\beta}\right|_{\delta\bm{\tau}_i=0},\quad \varepsilon_{\mathrm{s}xx}=1+\frac{1}{\epsilon_0}\left.\frac{\Delta P_\alpha}{E_\beta}\right|_{\bm{F}=0}\quad (\alpha,\beta=x,y,z),\label{eq:epsilon2}
\end{equation}
where
\(\Delta P_\alpha=P_\alpha(\bm{E}=E_\beta\hat{\mathbf{e_\beta}})-P_\alpha(\bm{E}=\bm0)\).
In addition, through the forces \(\bm{F}\) on atoms without the atomic
relaxation, the Born effective charge tensor \(Z^*\) of an atom was
also evaluated from
\begin{equation}
Z_{\alpha\beta}^*=\left.\frac{\partial F_\alpha}{\partial E_\beta}\right|_{\delta\bm\tau_i=\bm0}=\left.\frac{F_\alpha(\bm{E}=E_\beta\hat{\mathbf{e_\beta}})-F_\alpha(\bm{E}=\bm0)}{E_\beta}\right|_{\delta\bm\tau_i=\bm0}\quad (\alpha,\beta=x,y,z).\label{eq:Z^*}
\end{equation}
We then evaluated \(\varepsilon_{\infty xx}\),
\(\varepsilon_{\mathrm{s}xx}\) and \(Z_{xx}^*\). Here, we drop the subscripts and denote them as
\(\varepsilon_{\infty}\), \(\varepsilon_\mathrm{s}\) and \(Z^*\). The
finite electric field was considered after SCF convergence under the
zero field of each SCF loop, i.e., its effective potential was added
after the first SCF convergence. The matrix elements of the effective
potential was calculated from the Hamiltonian matrix in the previous SCF
iteration and the overlap matrices. The overlap matrices were prepared
at the beginning of each SCF loop. We noted that the forces converged when
the magnitudes were less than \(1\times10^{-5}\) Hartree/Bohr. This is
because a \(Z^*\) value of 1 yields a force of approximately \(2\times10^{-4}\)
Hartree/Bohr under an electric field of 0.1 GV/m, and highly precise
evaluation of the forces is essential to describe the electronic
structures under electric fields with
\(\varepsilon_\mathrm{s}\) and \(Z^*\). In the LCPAO method,
the so-called egg box effect leads to numerical errors arising from the real
space grid, and it is critical in evaluating the forces, dipole moment,
and stress tensor. The egg box effect can be resolved by grid cell
sampling and iterative evaluation of such quantities with 
grid shifting and frozen density matrix for the LCPAO ~{[}10{]}.
The density gradient in the GGA is sensitive to the egg box effect,
especially since it is estimated from the electronic density on the
real space grid. Therefore, we applied grid cell sampling to
evaluate the forces precisely by introducing a fine real space grid of
at least \(100\times100\times100\) after SCF convergence in the GGA
cases. Another workaround is to consider a stronger electric field to
enhance contributions from the electric field to the forces, but it
sometimes violates the limitation of the electric field strength due to
the Zener breakdown~{[}5{]}, and we did not treat the workaround in this work. To
evaluate \(\varepsilon_\mathrm{s}\), the atomic
relaxation needs to be considered. However, the computational cost of the grid cell sampling
with such fine grid was significantly high, because at
least eight evaluations are required (e.g.~\(2\times2\times2\) grid
shifts). Therefore, we used the \text{``}one-shot\text{''} grid cell
sampling, i.e., the forces based on the fine grid were evaluated from
the density matrix obtained from the rough grid without the grid shifts.

\hypertarget{results-and-discussion}{%
\section{Results and discussion}\label{results-and-discussion}}

\begin{table}[tbp]
\centering

\caption{Electronic dielectric constants of semiconducting materials.
Theoretical values of Ref.~{[}20{]} were obtained from density functional perturbation theory with
GGA, while values of Ref.~{[}5{]} and Ref.~{[}9{]} were calculated from
DFT with LDA and GGA, respectively, and the finite electric
field methods based on MTP. Our calculated values under three types
of conditions for the exchange correlation functional and the quality of
PAOs are shown. \label{tbl:eps_infty}}

\begin{tabular}{@{}rrrrrr@{}}
\hline

System & LDA;\emph{standard} & LDA;\emph{precise} & GGA;\emph{precise} & Ref.
(theor.) & Ref. (expt.) \\\hline

C & 5.61 & 5.67 & 5.67 & 5.9~{[}20{]} & 5.7~{[}21{]} \\
Si & 12.17 & 12.74 & 12.36 & 12.9~{[}20{]} & 11.6~{[}21{]} \\
AlP & 8.05 & 8.16 & 7.89 & 8.1~{[}5,20{]},
7.84~{[}9{]} & 7.4~{[}21{]} \\
AlAs & 8.84 & 9.05 & 8.85 & 9.6~{[}5{]}, 8.80~{[}9{]},
9.3~{[}20{]} & 8.16~{[}21{]} \\
GaP & 10.33 & 10.45 & 9.94 & 9.4~{[}5{]},
10.4~{[}20{]} & 8.8~{[}21{]} \\
GaAs & 13.84 & 14.42 & 13.16 & 11.9~{[}5{]},
13.7~{[}20{]} & 10.86~{[}21{]} \\
AlSb & 10.84 & 11.14 & 10.90 & 11.45~{[}9{]},
11.5~{[}20{]} & 9.88~{[}21{]} \\
InP & 11.07 & 10.91 & 10.08 & 11~{[}20{]} & 9.9~{[}21{]} \\
ZnS & 5.93 & 6.12 & 5.77 & 5.9~{[}20{]} & 5.1~{[}21{]} \\
ZnSe & 7.15 & 7.41 & 6.90 & 7.2~{[}20{]} & 5.9~{[}21{]} \\
ZnTe & 8.35 & 8.86 & 8.24 & 8.9~{[}20{]} & 6.9~{[}21{]} \\
CdSe & 7.64 & 8.21 & 7.21 & 7.6~{[}20{]} & 6.2~{[}21{]} \\
CdTe & 8.09 & 8.75 & 7.89 & 8.6~{[}20{]} & 7.1~{[}21{]} \\
MgO & 3.09 & 3.18 & 3.14 & 3.1~{[}20{]} & 3.1~{[}21{]} \\

\hline
\end{tabular}
\end{table}

To validate our formalism, we calculated the electronic dielectric constants
\(\varepsilon_{\infty}\) of III-V, II-VI semiconductors and group IV
insulators. First, we calculated \(\varepsilon_\infty\) with LDA to
compare our values to those from the plane wave formalism based on the
MTP by Souza \emph{et al.}~{[}5{]}. Our calculated values are shown in
\cref{tbl:eps_infty}. Our values of the AlP and AlAs based on the
\emph{precise} PAO basis set coincide with their values, while the ones
of the GaP and GaAs have more than 10\% and 20\% errors. Although our
LCPAO formalism is equivalent to Souza \emph{et al.}'s formalism, this
difference might arise from different conditions such as
pseudopotentials. Indeed, our measured values are in good agreement with
recently reported values under GGA based on the
density functional perturbation theory. Next, we calculated
\(\varepsilon_\infty\) under the GGA, and the values were improved
(i.e.~closer to experimental values) for all the systems compared to 
LDA. We confirmed that a moderately large number of k-points are necessary to
obtain the converged values as reported by Souza et al.~{[}5{]}. We also
investigated the PAO quality dependence, and the values for the
\emph{standard} basis sets are smaller than those for the \emph{precise}
ones, except the InP. The \emph{standard} basis sets yield PAOs around
atom centers in a more narrow domain than the \emph{precise} basis sets.
A more localized description of electronic states through the
\emph{standard} basis sets can constrain the electronic distribution more
strongly, and suppress \(\varepsilon_\infty\). These results
demonstrated that our LCPAO implementation of the effective potential for
the finite electric field was accurate.

To investigate the influence of the SOI, we focused on ZnTe and CdTe
including tellurium, which is the heaviest element in our systems. When considering the SOI with the \emph{precise} basis set
under GGA, the \(\varepsilon_\infty\) of ZnTe and CdTe was 8.37
and 8.10, and increased by 0.13 and 0.21, respectively. The strong SOI
can drastically change the electronic structure so that these increases
occur. In fact, the band gap of ZnTe and CdTe changed from 1.73 eV
and 1.04 eV to 1.60 eV and 0.77 eV after including the SOI,
respectively (see also Appendix A for the relationship between the band
gap and \(\varepsilon_\infty\)). The \(\varepsilon_\infty\) of AlSb also
increased by 0.11, and it became 11.01. Our results illustrated that for
systems including heavy-elements, the SOI effect on 
\(\varepsilon_\infty\) should be considered.

\begin{table}[tbp]
\centering

\caption{Born effective charges of semiconducting materials. Theoretical
values of Refs.~{[}22--24{]} were obtained from DFT with LDA and
the finite difference method based on the MTP, while those of
Refs.~{[}5,8{]} and Ref.~{[}9{]} were calculated from DFT with
LDA and GGA, respectively, and the finite electric field methods
based on the MTP. Our calculated values under two types of conditions
for the exchange correlation functional and the quality of PAOs are
shown. The value of Ref.~{[}24{]} is that of the anion. \label{tbl:Z}}

\begin{tabular}{@{}rrrrr@{}}
\hline

System & LDA;\emph{precise} & GGA;\emph{precise} & Ref. (theor.) & Ref.
(expt.) \\\hline

AlP & 2.19 & 2.20 & 2.24~{[}5{]}, 2.23~{[}9{]},
2.20~{[}22{]} & 2.15~{[}25{]} \\
AlAs & 2.12 & 2.09 & 2.14~{[}5{]}, 2.110~{[}23{]}, 2.17~{[}9{]},
2.12~{[}22{]} & 2.20~{[}26{]} \\
GaP & 2.05 & 2.11 & 2.10~{[}5{]}, 2.06~{[}22{]} & 2.16~{[}27{]} \\
GaAs & 2.09 & 2.10 & 2.00~{[}5{]}, 2.18~{[}23{]},
2.08~{[}22{]} & 2.18~{[}26{]} \\
AlSb & 1.81 & 1.75 & 1.83~{[}9{]}, 1.81~{[}22{]} & 1.93~{[}28{]} \\
InP & 2.44 & 2.48 & 2.50~{[}22{]} & 2.55~{[}29{]} \\
ZnS & 1.87 & 1.89 & -1.99~{[}24{]} & 2.15~{[}29{]} \\
ZnSe & 1.95 & 1.95 & 2.12~{[}30{]} & 2.03~{[}29{]} \\
ZnTe & 1.87 & 1.86 &  & 2.00~{[}29{]} \\
CdSe & 2.19 & 2.17 &  & 2.25~{[}29{]} \\
CdTe & 2.09 & 2.08 &  & 2.35~{[}29{]} \\
MgO & 1.98 & 1.98 & 1.96~{[}8{]} & 1.77~{[}29{]} \\

\hline
\end{tabular}
\end{table}

As shown in \Cref{tbl:Z}, we also calculated the Born effective charges
\(Z^*\) of cations except for the group IV systems to validate our
formalism of the forces on atoms under electric fields. Our
calculated results were robust with respect to exchange correlation
functionals, and each \(Z^*\) under the LDA is approximately equal to that
under GGA. Our values were in good agreement with not only the
theoretical values, but also experimental values in previous studies.
The mean absolute error of \(Z^*\) for our values and the
experimental values with respect to the systems was \textasciitilde6\%.
Therefore, we concluded that our formalism of forces under finite
electric fields was also correct. In including the SOI, however, only
negligible changes appeared. Although a fine real space grid allowed us
to get converged values, we confirmed that even a rough real space grid
could provide good values, which suggests that the computational cost
can be reduced in calculating \(Z^*\).

\begin{table}[tbp]
\centering

\caption{Static dielectric constants of semiconducting materials.
Theoretical values of Ref.~{[}5,8{]} and Ref.~{[}9{]} were calculated
from DFT with LDA and GGA, respectively, and the finite
electric field methods based on MTP. Our calculated values under two
types of conditions for the exchange correlation functional and the
quality of PAOs are shown. \label{tbl:eps_static}}

\begin{tabular}{@{}rrrrr@{}}
\hline

System & LDA;\emph{precise} & GGA;\emph{precise} & Ref. (theor.) & Ref.
(expt.) \\\hline

AlP & 10.26 & 9.94 & 10.2~{[}5{]}, 10.26~{[}9{]} & 9.6~{[}21{]} \\
AlAs & 10.90 & 10.56 & 11.5~{[}5{]}, 11.05~{[}9{]} & 10.6~{[}21{]} \\
GaP & 11.96 & 11.69 & 11.2~{[}5{]} & 11.0~{[}21{]} \\
GaAs & 19.28 & 14.38 & 13.5~{[}5{]} & 12.90~{[}21{]} \\
MgO & 7.92 & 8.79 & 7.93~{[}8{]} & 9.8~{[}21{]} \\

\hline
\end{tabular}
\end{table}

Then, we evaluated the atomic relaxation under finite electric fields
through the static dielectric constants \(\varepsilon_\mathrm{s}\). As
shown in \Cref{tbl:eps_static}, the values based on LDA and GGA
were successfully reproduced, except for GaAs. The LDA value of GaAs was
overestimated compared to the results of previous studies, because our
calculated band gap of 0.25 eV was underestimated compared to the
experimental value of 1.579 eV~{[}31{]}, which made more delocalized
pictures of electrons to give the overestimation. The GGA values were
closer to the experimental ones than those of LDA, and the AlP, AlAs, and GaP
values had errors within 7\%, while the errors for GaAs and MgO 
were within \textasciitilde30\%. Although we reproduced the LDA value of
MgO reported in the previous theoretical study~{[}8{]}, there was a
mismatch of the MgO values between the experimental and theoretical
results. With an experimental lattice constant of \(4.212 \AA\)~{[}32{]}
for its conventional cell, the \(\varepsilon_\mathrm{s}\) of MgO became
9.56, i.e., it approached the experimental \(\varepsilon_\infty\) of
9.8~{[}21{]}, although our lattice constant was 1\% smaller compared to
the experimental value. In the atomic relaxation, the precision of the
forces on atoms is of importance, and even minor errors in the
magnitude of \(10^{-4}\) Hartree/Bohr can reflect large errors in
\(\varepsilon_\mathrm{s}\). As a result, it is suggested that not only a
fine k-space grid, but also a fine real space grid is required to
describe the movements of atoms in solids under a finite electric field.
Although the computational cost of the ordinary grid cell sampling with
a fine real space grid was significantly high to continue relaxation, we
confirmed that the \text{``}one shot\text{''} grid cell sampling was
efficient and effective in avoiding the numerical errors induced by the
real space grid or density gradient. Finally, it was proved that our
formalism for the forces was correct, and that it can be applied to both atomic relaxation and molecular dynamics under finite electric
fields.

\hypertarget{conclusion}{%
\section{Conclusion}\label{conclusion}}

We developed a first-principle LCPAO scheme of a finite electric field method
based on MTP to yield self-consistent solutions, including the forces
on atoms in the DFT framework. Moreover, we investigated the PAO dependence and exchange correlation dependence by performing systematic calculations.
Our implementation successfully reproduced the electronic and static
dielectric constants and Born effective charges of III-V and II-VI
semiconductors and group IV insulating materials. We confirmed that the
implementation for the SOI is necessary for systems including heavy
elements. Our LCPAO implementation is expected to be advantageous over
conventional formalisms such as the plane wave method in
terms of efficiency and extensibility, especially for large scale
calculations modeling realistic systems through the \(O(N)\) method, and
the atomic or orbital decomposition of the effective potential. Our formalism will expand the problem size such that
complicated systems such as devices can be considered, and the
molecular dynamics of complicated systems such as batteries under a
finite electric field can be realized with a lower computational cost.

\hypertarget{acknowledgement}{%
\section*{Acknowledgement}\label{acknowledgement}}
\addcontentsline{toc}{section}{Acknowledgement}

This work was supported by JSPS KAKENHI Grants Numbers JP18J21257,
JP18H04481, JP19H02554, JP20K15115. The computation in this work has
been conducted using the facilities of the Supercomputer Center, the
Institute for Solid State Physics, the University of Tokyo. The
computation was conducted using the computer resource offered under
the category of General Projects by Research Institute for Information
Technology, Kyushu University. This work used computational resources of
the supercomputer Fugaku provided by the RIKEN Center for Computational
Science through the HPCI System Research Project (Project ID: hp210305).
We would like to thank Editage (www.editage.com) for English language editing.

\hypertarget{appendix-a-correlation-between-the-varepsilon_infty-and-the-band-gap}{%
\section*{\texorpdfstring{Appendix A: Correlation between the
\(\varepsilon_\infty\) and the band
gap}{Appendix A: Correlation between the \textbackslash varepsilon\_\textbackslash infty and the band gap}}\label{appendix-a-correlation-between-the-varepsilon_infty-and-the-band-gap}}
\addcontentsline{toc}{section}{Appendix A: Correlation between
\(\varepsilon_\infty\) and the band gap}

\begin{figure}
\hypertarget{fig:eps_infty_vs_gap}{%
\centering
\includegraphics{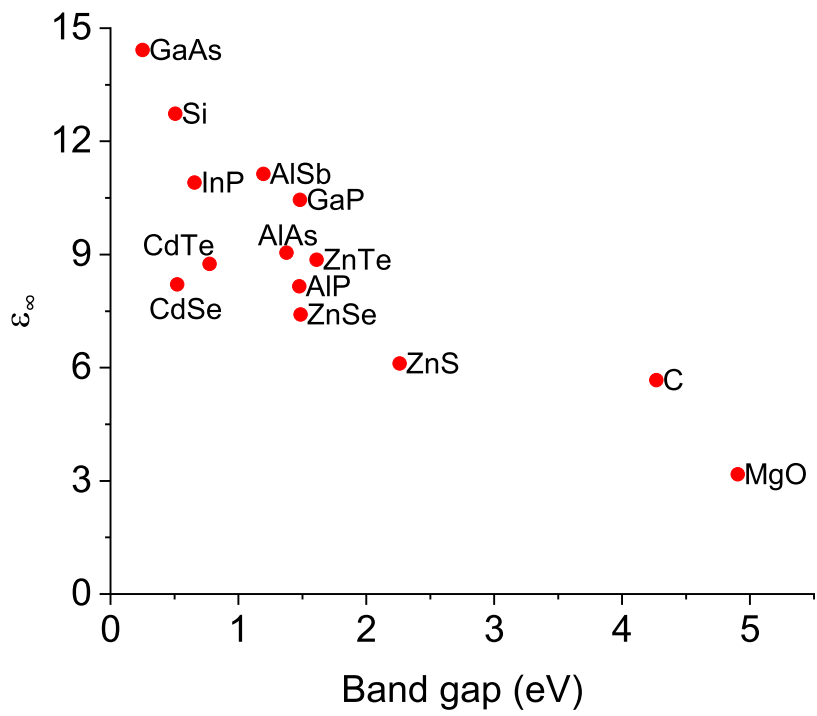}
\caption{\(\varepsilon_\infty\) versus the band gap calculated with the LDA functional and \emph{precise} basis
sets.}\label{fig:eps_infty_vs_gap}
}
\end{figure}

\Cref{fig:eps_infty_vs_gap} shows that there was a trade-off between the
calculated \(\varepsilon_\infty\) and band gap. This trend coincides
with the picture that the electron distribution is more localized for the
wider band gap.

\hypertarget{references}{%
\section*{References}\label{references}}
\addcontentsline{toc}{section}{References}

\hypertarget{refs}{}
\begin{CSLReferences}{0}{0}
\leavevmode\hypertarget{ref-Ehrenreich1959}{}%
\CSLLeftMargin{{[}1{]} }
\CSLRightInline{H. Ehrenreich and M. H. Cohen, Phys. Rev. \textbf{115},
786 (1959).}

\leavevmode\hypertarget{ref-Sham1969}{}%
\CSLLeftMargin{{[}2{]} }
\CSLRightInline{L. J. Sham, Phys. Rev. \textbf{188}, 1431 (1969).}

\leavevmode\hypertarget{ref-Pick1970}{}%
\CSLLeftMargin{{[}3{]} }
\CSLRightInline{R. M. Pick, M. H. Cohen, and R. M. Martin, Phys. Rev. B
\textbf{1}, 910 (1970).}

\leavevmode\hypertarget{ref-King-Smith1993}{}%
\CSLLeftMargin{{[}4{]} }
\CSLRightInline{R. D. King-Smith and D. Vanderbilt, Phys. Rev. B
\textbf{47}, 1651 (1993).}

\leavevmode\hypertarget{ref-Souza2002}{}%
\CSLLeftMargin{{[}5{]} }
\CSLRightInline{I. Souza, J. Íñiguez, and D. Vanderbilt, Phys. Rev.
Lett. \textbf{89}, 9 (2002).}

\leavevmode\hypertarget{ref-Nunes1994}{}%
\CSLLeftMargin{{[}6{]} }
\CSLRightInline{R. W. Nunes and D. Vanderbilt, Phys. Rev. Lett.
\textbf{73}, 712 (1994).}

\leavevmode\hypertarget{ref-Fernandez1998}{}%
\CSLLeftMargin{{[}7{]} }
\CSLRightInline{P. Fernández, A. Dal Corso, and A. Baldereschi, Phys.
Rev. B - Condens. Matter Mater. Phys. \textbf{58}, R7480 (1998).}

\leavevmode\hypertarget{ref-Umari2002}{}%
\CSLLeftMargin{{[}8{]} }
\CSLRightInline{P. Umari and A. Pasquarello, Phys. Rev. Lett.
\textbf{89}, 1 (2002).}

\leavevmode\hypertarget{ref-Zwanziger2012}{}%
\CSLLeftMargin{{[}9{]} }
\CSLRightInline{J. W. Zwanziger, J. Galbraith, Y. Kipouros, M. Torrent,
M. Giantomassi, and X. Gonze, Comput. Mater. Sci. \textbf{58}, 113
(2012).}

\leavevmode\hypertarget{ref-Soler2002}{}%
\CSLLeftMargin{{[}10{]} }
\CSLRightInline{J. M. Soler, E. Artacho, J. D. Gale, A. García, J.
Junquera, P. Ordejón, and D. Sánchez-Portal, J. Phys. Condens. Matter
\textbf{14}, 2745 (2002).}

\leavevmode\hypertarget{ref-Souza2004}{}%
\CSLLeftMargin{{[}11{]} }
\CSLRightInline{I. Souza, J. Íñiguez, and D. Vanderbilt, Phys. Rev. B -
Condens. Matter Mater. Phys. \textbf{69}, 1 (2004).}

\leavevmode\hypertarget{ref-Artacho2008}{}%
\CSLLeftMargin{{[}12{]} }
\CSLRightInline{E. Artacho, E. Anglada, O. Diéguez, J. D. Gale, A.
García, J. Junquera, R. M. Martin, P. Ordejón, J. M. Pruneda, D.
Sánchez-Portal, and J. M. Soler, J. Phys. Condens. Matter \textbf{20},
064208 (2008).}

\leavevmode\hypertarget{ref-Berkaine2011}{}%
\CSLLeftMargin{{[}13{]} }
\CSLRightInline{N. Berkaïne, E. Orhan, O. Masson, P. Thomas, and J.
Junquera, Phys. Rev. B - Condens. Matter Mater. Phys. \textbf{83},
245205 (2011).}

\leavevmode\hypertarget{ref-Xiang2006}{}%
\CSLLeftMargin{{[}14{]} }
\CSLRightInline{H. J. Xiang, J. Yang, J. G. Hou, and Q. Zhu, Phys. Rev.
Lett. \textbf{97}, 1 (2006).}

\leavevmode\hypertarget{ref-Ozaki2003}{}%
\CSLLeftMargin{{[}15{]} }
\CSLRightInline{T. Ozaki, Phys. Rev. B - Condens. Matter Mater. Phys.
\textbf{67}, 1 (2003).}

\leavevmode\hypertarget{ref-Ozaki2004}{}%
\CSLLeftMargin{{[}16{]} }
\CSLRightInline{T. Ozaki and H. Kino, Phys. Rev. B - Condens. Matter
Mater. Phys. \textbf{69}, 1 (2004).}

\leavevmode\hypertarget{ref-Ozaki2005}{}%
\CSLLeftMargin{{[}17{]} }
\CSLRightInline{T. Ozaki and H. Kino, Phys. Rev. B - Condens. Matter
Mater. Phys. \textbf{72}, 1 (2005).}

\leavevmode\hypertarget{ref-Ceperley1980}{}%
\CSLLeftMargin{{[}18{]} }
\CSLRightInline{D. M. Ceperley and B. J. Alder, Phys. Rev. Lett.
\textbf{45}, 566 (1980).}

\leavevmode\hypertarget{ref-Perdew1996}{}%
\CSLLeftMargin{{[}19{]} }
\CSLRightInline{J. P. Perdew, K. Burke, and M. Ernzerhof, Phys. Rev.
Lett. \textbf{77}, 3865 (1996).}

\leavevmode\hypertarget{ref-Hinuma2017}{}%
\CSLLeftMargin{{[}20{]} }
\CSLRightInline{Y. Hinuma, Y. Kumagai, I. Tanaka, and F. Oba, Phys. Rev.
B \textbf{95}, (2017).}

\leavevmode\hypertarget{ref-doi:https:ux2fux2fdoi.orgux2f10.1002ux2f0470090340.ch10}{}%
\CSLLeftMargin{{[}21{]} }
\CSLRightInline{in \emph{Properties of Group‐IV, III‐v and II‐VI
Semiconductors} (John Wiley \& Sons, Ltd, 2005), pp. 211--281.}

\leavevmode\hypertarget{ref-Benyahia2017}{}%
\CSLLeftMargin{{[}22{]} }
\CSLRightInline{N. Benyahia, A. Zaoui, D. Madouri, and M. Ferhat, J.
Appl. Phys. \textbf{121}, (2017).}

\leavevmode\hypertarget{ref-Wang2007}{}%
\CSLLeftMargin{{[}23{]} }
\CSLRightInline{X. Wang and D. Vanderbilt, Phys. Rev. B - Condens.
Matter Mater. Phys. \textbf{75}, 1 (2007).}

\leavevmode\hypertarget{ref-DalCorso1994}{}%
\CSLLeftMargin{{[}24{]} }
\CSLRightInline{A. Dal Corso, M. Posternak, R. Resta, and A.
Baldereschi, Phys. Rev. B \textbf{50}, 10715 (1994).}

\leavevmode\hypertarget{ref-Lakel2015}{}%
\CSLLeftMargin{{[}25{]} }
\CSLRightInline{S. Lakel, F. Okbi, M. Ibrir, and K. Almi, AIP Conf.
Proc. \textbf{1653}, (2015).}

\leavevmode\hypertarget{ref-Spencer1994}{}%
\CSLLeftMargin{{[}26{]} }
\CSLRightInline{G. S. Spencer, A. C. Ho, J. Menéndez, R. Droopad, H.
Fathollahnejad, and G. N. Maracas, Phys. Rev. B \textbf{50}, 14125
(1994).}

\leavevmode\hypertarget{ref-Samarasingha2021}{}%
\CSLLeftMargin{{[}27{]} }
\CSLRightInline{N. S. Samarasingha and S. Zollner, J. Vac. Sci. Technol.
B \textbf{39}, 052201 (2021).}

\leavevmode\hypertarget{ref-Ves1986}{}%
\CSLLeftMargin{{[}28{]} }
\CSLRightInline{S. Ves, K. Strössner, and M. Cardona, Solid State
Commun. \textbf{57}, 483 (1986).}

\leavevmode\hypertarget{ref-Lucovsky1971}{}%
\CSLLeftMargin{{[}29{]} }
\CSLRightInline{G. Lucovsky, R. M. Martin, and E. Burstein, Phys. Rev. B
\textbf{4}, 1367 (1971).}

\leavevmode\hypertarget{ref-Wang2014}{}%
\CSLLeftMargin{{[}30{]} }
\CSLRightInline{D. Wang, X. Zhang, B. Li, L. Liu, and D. Z. Shen, AIP
Adv. \textbf{4}, (2014).}

\leavevmode\hypertarget{ref-haynes2016crc}{}%
\CSLLeftMargin{{[}31{]} }
\CSLRightInline{W. M. Haynes, \emph{CRC Handbook of Chemistry and
Physics} (CRC Press, 2016).}

\leavevmode\hypertarget{ref-Sun1999}{}%
\CSLLeftMargin{{[}32{]} }
\CSLRightInline{D. Sun, H. Enoki, F. Gingl, and E. Akiba, J. Alloys
Compd. \textbf{285}, 279 (1999).}

\end{CSLReferences}

\end{document}